  \documentclass[a4paper,11pt]{article}
\usepackage{amsmath,graphicx,color,mathrsfs,multicol}
 
\usepackage{verbatim}
 
\input epsf.sty

\newcommand{\com}[1]{}

\def\circa#1{\,\raise.3ex\hbox{$#1$\kern-.75em\lower1ex\hbox{$\sim$}}\,}
\makeatletter

\def\art{\@ifnextchar[{\eart}{\oart}}
\def\eart[#1]#2#3#4#5#6{{\rm #2}, {\em #3  #4} {\rm (#6) #5} ({\em #1})}
\def\hepart[#1]#2{{\rm #2, \em#1}}
\newcommand{\oart}[5]{{\rm #1}, {\em #2  #3} {\rm (#5) #4}}

\newcounter{alphaequation}[equation]
\def\thealphaequation{\theequation\hbox to
0.6em{\hfil\alph{alphaequation}\hfil}}
\def\eqnsystem#1{
\def\@eqnnum{{\rm (\thealphaequation)}}
\def\@@eqncr{\let\@tempa\relax \ifcase\@eqcnt \def\@tempa{& & &} \or
  \def\@tempa{& &}\or \def\@tempa{&}\fi\@tempa
  \if@eqnsw\@eqnnum\refstepcounter{alphaequation}\fi
\global\@eqnswtrue\global\@eqcnt=0\cr}
\refstepcounter{equation} \let\@currentlabel\theequation \def\@tempb{#1}
\ifx\@tempb\empty\else\label{#1}\fi
\refstepcounter{alphaequation}
\let\@currentlabel\thealphaequation
\global\@eqnswtrue\global\@eqcnt=0 \tabskip\@centering\let\\=\@eqncr
$$\halign to \displaywidth\bgroup \@eqnsel\hskip\@centering
$\displaystyle\tabskip\z@{##}$&\global\@eqcnt\@ne
\hskip2\arraycolsep\hfil${##}$\hfil& \global\@eqcnt\tw@\hskip2\arraycolsep
$\displaystyle\tabskip\z@{##}$\hfil
\tabskip\@centering&\llap{##}\tabskip\z@\cr}
\def\endeqnsystem{\@@eqncr\egroup$$\global\@ignoretrue} \makeatother

\usepackage{epsfig,multicol}
\usepackage{amsmath,amsfonts,amssymb}
\input epsf.sty

\usepackage{graphicx}
\usepackage[body={17.5cm, 22cm},right=2.2cm]{geometry}
\usepackage{amssymb}
\usepackage{amsmath}

\allowdisplaybreaks[1]

\def\o+{\oplus}

\def\beqa{\begin{eqnarray}}
\def\eeqa{\end{eqnarray}}

\newcommand{\be}{\begin{equation}}
\newcommand{\ee}{\end{equation}}
\newcommand{\bea}{\begin{eqnarray}}
\newcommand{\eea}{\end{eqnarray}}
\newcommand{\bg}{\begin{gather}}
\newcommand{\eg}{\end{gather}}
\newcommand{\bseq}{\begin{subequations}}
\newcommand{\eseq}{\end{subequations}}

\newcommand{\cotan}{{\rm cotan}}

\def\be{\begin{equation}}
\def\EQ{\begin{equation}}
\def\ee{\end{equation}}
\def\EN{\end{equation}}
\def\bea{\begin{eqnarray}}
\def\beq{\begin{eqnarray}}
\def\ba{\begin{eqnarray}}
\def\eea{\end{eqnarray}}
\def\ena{\end{eqnarray}}
\def\eeq{\end{eqnarray}}
\def\ea{\end{eqnarray}}
\def\bei{\begin{itemize}}
\def\eei{\end{itemize}}
\def\bee{\begin{enumerate}}
\def\eee{\end{enumerate}}

\def\lx{\left}
\def\rx{\right}

\def\a{\alpha}

\begin{document}
\vspace{1.5cm}

\begin{center}
{\LARGE \bf Induced gravity 
  on intersecting brane-worlds
  
  \vskip 0.4cm
  
  {\Large Part I: Maximally symmetric 
  solutions}
}\\[1cm]

{
{\large\bf Olindo Corradini$^{a\,}$,\,  Kazuya Koyama$^{b\,}$
\,and\, Gianmassimo Tasinato$^{c\,}$
}
}
\\[7mm]
{\it $^a$ Dipartimento di Fisica, Universit\`a di Bologna and INFN Sezione
  di Bologna\\ Via Irnerio, 46 - Bologna I-40126, Italy} \\ 
 {E-mail}:  corradini@bo.infn.it\\[3mm]
{\it $^b$ Institute of Cosmology and Gravitation, 
University of Portsmouth  \\ Portsmouth PO1 2EG, UK 
}\\ {E-mail}: Kazuya.Koyama@port.ac.uk  \\[3mm]
{\it $^c$ Instituto de Fisica Teorica, UAM/CSIC  \\ Facultad de Ciencias C-XVI, C.U. Cantoblanco,
E-28049-Madrid, Spain} \\
 {E-mail}:  gianmassimo.tasinato@uam.es
\\[1cm]
\vspace{-0.3cm}

\vspace{1cm}

{\large\bf Abstract}

\end{center}
\begin{quote}

{
We explore models of intersecting brane-worlds with induced gravity terms
on codimension one branes and on their intersection. Maximally symmetric solutions 
for the branes and the intersection are found. We find new self-accelerating 
solutions. In a 6d spacetime, the solutions realize the see-saw modification of 
gravity where the UV scale of the modification to 4d gravity is determined by 
6d Planck scale given by $M_6 \sim 10^{-3}$eV and the IR scale of the modification 
is determined by $M_6^2/M_4 \sim H_0 \sim 10^{-42}$ GeV where $H_0$ is present-day 
Hubble scale. We find that it is increasingly difficult to construct phenomenologically 
viable models in higher-dimensional spacetime due to the necessity to
have the lower value for the fundamental Planck scale to realize the late time acceleration. 
It is found that the system also admits self-tuning solutions where the tension 
at the intersection does not change the geometry of the intersection. 
The induced gravity terms can avoid the necessity to compactify the 
extra dimensions. Finally, we discuss the possibility to have ordinary
matter at the intersection, without introducing any regularisation,
using the induced gravity terms.

}

\end{quote}

\newpage
\tableofcontents

\bigskip

\bigskip

\bigskip

\section{Introduction}\label{intro}

Brane-world models with large distance modification
of Einstein gravity are invoked in scenarios that aim 
to geometrically describe present day 
acceleration, without introducing dark energy
\cite{dgp1,deffayetcosmology,deffayetdvali} (for a review see 
\cite{KK-review}). 
A celebrated example is the DGP model in a 5d spacetime~\cite{dgp1}.  
The brane action includes a quantum-induced Einstein-Hilbert that 
recovers 4d gravity on small scales. 
This model realizes a so-called  {\it self-accelerating} solution that 
features a 4d de Sitter phase even though the 3-brane is
completely empty. However, so far, only codimension-one examples of such
solutions have been proposed and these backgrounds are known to
suffer from  ghost instabilities~\cite{Luty:2003vm}. An interesting
possibility would thus be to look for other such solutions in 
higher codimension set-ups,  that might lead to ghost free models.    
 
 \smallskip
 
On the other hand, it is known 
that the presence of more than one extra dimensions might offer better
possibilities for the solution of the (particle physics) cosmological
constant problem. In fact, codimension-two brane-worlds with
finite~\cite{firstattemptscod2,carroll,sled,burgess1} and
infinite~\cite{Gregory:2000jc,dvaligabadadze} volume 
extra space have been extensively studied in the past few years
towards a solution of the cosmological constant problem.~\footnote{See
\cite{cod1cc} for attempts in this 
direction in the codimension one case.} 
A common feature of all these models is that 
the energy momentum tensor on a codimension two brane only controls 
the {\it global} properties of the background geometry without affecting the 
local structure of the brane itself which remains flat.
A realization of this idea is called the  
{\it selftuning} solution of the cosmological constant problem.   
Unfortunately, in the finite volume case, consistency conditions associated 
to the quantisation condition of a background flux which is introduced  
to compactify the extra-dimensional space are shown to be problematic and the 
selftuning  mechanism in its simplest form does not work~\cite{Garriga:2004tq}. 
More in general, another delicate feature of codimension two brane models (both 
with finite- and infinite-volume) is related to the fact 
that only energy momentum tensor of the form of  {\it pure tension} can be 
accommodated on a thin radially-symmetric codimension 
 two brane, 
to ensure the regularity of the background  metric at the brane position 
\cite{puretension1}.  This last problem can be
ameliorated by including higher order Gauss-Bonnet terms for gravity
in the bulk \cite{puretension2}; however other difficulties arise when
one tries to embed an isotropic and homogeneous fluid on such a
radially-symmetric purely-conical codimension two brane~\cite{greeks}.
 This suggests the necessity of regularising the brane and allowing it
to acquire some structure. For the finite volume scenario, this approach was 
started by \cite{pst} that shows how normal gravity can be recovered in this
context,
 and it is further developed by various other works
\cite{alltherestregularization}~\footnote{See also \cite{charmousis}
  for articles discussing gravity in higher codimensions.}.  
The regularisation of the co-dimension two brane with 
induced gravity has been also studied intensively \cite{6d-induced1, 6d-induced2}. 

\smallskip

In this paper, we take a different route to tackle this
regularisation problem. We consider a system in which the brane
is a pure codimension two object, without an internal structure,
but with induced gravity terms on it. This 
system, as we already mentioned at the beginning, is by itself interesting to 
investigate, for finding new self-accelerating configurations.
The induced gravity terms can avoid the necessity to compactify the higher-dimensional 
space, by providing a mechanism to obtain
four dimensional gravity in the relevant regimes with infinite volume 
extra-dimensions. Consequently the bulk gauge field is no more needed, and 
this suggests to re-consider the self-tuning mechanism
in this context.

The model that we consider is a codimension
two brane, that lies at the intersection of two
codimension one branes embedded in a six dimensional
space. This system was studied in the context of standard
gravity~\cite{int-standard} 
(cosmological properties were studied in \cite{Cline:1999tq}),
and Gauss-Bonnet gravity \cite{leetasinato} elaborating an idea
developed in~\cite{lee2}.~\footnote{ 
Intersecting brane models are very important
in string theory, for the possibility to build Standard-like
models with chiral matter  \cite{strings}.} The latter was generalised
to higher codimensional models in~\cite{ignaciomulti}. 
Models with a generic angle 
between two intersecting branes were first considered in
\cite{Csaki:1999mz} and then further generalised into the so-called
Origami-world in~\cite{origami}. 
In our case, we consider a situation in which induced gravity terms
are allowed on the codimension one branes and also  
at the intersection. We study various maximally symmetric 
configurations of branes and  
explore configurations that exhibit selftuning or self-accelerating 
properties. Let us also point out that the presence of induced Einstein-Hilbert 
terms on intersecting codimension-one branes might allow for generic localised 
matter on the intersection and thus allow for generic FRW cosmology.  
We discuss this possibility in conclusions, briefly
summarising the results of a companion paper \cite{inpreparation}.

\smallskip

While we were preparing this work, a relevant paper \cite{newdvali}
appeared that studied a nested 4d brane in a 5d codimension brane
in a 6d spacetime by generalizing \cite{nestedgregory}.
 
\section{Maximally symmetric configurations of branes}\label{statisec}

We consider a system of two intersecting codimension one branes 
embedded in a six dimensional space-time. They intersect on a four dimensional 
codimension two brane where observers like us can be localised.
We take an Einstein-Hilbert action for gravity in the bulk  
and we allow for induced gravity terms on the codimension one branes,
as well as on the intersection. Besides gravity, we allow for 
a cosmological constant term in the bulk, $\Lambda_B$, and 
for additional fields localised on the branes described by general 
Lagrangians $L$'s. The  general action assumes the form 

\begin{eqnarray}\label{model}
S&=&\int_{\rm bulk} d^{6} x \sqrt{-g}\ \biggl(\frac{M_{6}^{4}}2 R-\Lambda_B\biggr)
+\sum_{i=1}^2 \int_{\Sigma_i} d^{5} x \sqrt{-g_{(i)}}\ \biggl(\frac{M_{5,i}^{3}}2
R_{(i)}+L_{(i)}\biggr)\nonumber\\    
&
+&\int_{\Sigma_{\cap}} d^{4} x \sqrt{-g_\cap}\ \biggl(\frac{M_{4}^{2}}2 R_\cap+L_\cap\biggr),
\end{eqnarray}
where $\Sigma_{\cap}
\equiv\bigcap_i \Sigma_i$ denotes a three-brane at the intersection between 
all codimension-one branes four-branes $\Sigma_i$. We can have different 
fundamental scales in the different regions of the space, $M_{6}$, $M_{5,i}$, 
and $M_{4}$. The induced gravity terms could be generated, as it was proposed
in the original model, by quantum corrections from matter loops on the brane.
It is also interesting to note that induced curvature terms appear quite generically 
in junction conditions of higher codimension branes when considering natural 
generalisations of Einstein gravity \cite{puretension2, charmousis} 
as well as in string theory compactifications 
\cite{antoniadis}, orientifold models and intersecting D-brane
models~\cite{Kohlprath:2003pu}.

\bigskip
 
In this paper, we focus on a configuration with {\it static}  
codimension one branes. They are characterised by tensions $\Lambda_1$ and
$\Lambda_2$, while the intersection has tension $\lambda$. 
The branes intersect with a generic angle, along the lines of 
Origami-world~\cite{origami}. The six dimensional bulk is characterised 
by a maximally symmetric geometry
\bea
ds^2 &=& A^2(t,z^1,z^2)\lx(\eta_{\mu\nu} dx^\mu dx^\nu +\delta_{k h}dz^k 
dz^h\rx), \nonumber\\  
A(t,z^1,z^2) &=& \frac{1}{Ht+k_i z^i}~.\label{bulkgengeo}
\eea
Also the codimension one branes and their intersection are characterised by 
maximally symmetric geometries. We are going to relate the Hubble parameters 
on the branes with the geometrical parameters that control our system.
  
The parameters $H$ and $k_i$ appearing  in the warp factor $A$ satisfy the 
following relation
\be \label{cobulk}
\frac{\Lambda_B}{10}\,=\,H^2-k_1^2-k_2^2,
\ee
in order to solve the Einstein equations in the bulk. 
The branes form a generic angle and are characterised by normal vectors
\bea
{\bf n}^{(1)} &=& (\sin \alpha_1, -\cos \alpha_1),\\
{\bf n}^{(2)} &=& (\sin \alpha_2, \cos \alpha_2),
\eea
in the two transverse space directions $z^i$, so that the angle
between the two branes is $\alpha \equiv \left(\alpha_1+\alpha_2 \right)$.
We refer to the brane $\Sigma_i$ as orthogonal to the vector ${\bf n}^{(i)}$. 
It is useful to define new coordinates 
\bea
\tilde z^k =  {\bf n}^{(k)} \cdot {\bf z}\quad {\rm with}\quad {\bf z}=(z^1,z^2), 
\eea
that allow us to define the brane positions as $\tilde{z}^i=0$.
Following \cite{origami}, we define new vectors $l_{(k)}$ as
\bea
{\bf l}_{(k)}\cdot {\bf n}^{(h)} = \delta_k{}^h
\quad \rightarrow\quad \left\{
\begin{array}{l}
{\bf l}_{(1)}\, = \,\frac{1}{\sin \alpha}(\cos\alpha_2,-\sin\alpha_2),\\
{\bf l}_{(2)} \,=\,
\,\frac{1}{\sin \alpha}(\cos\alpha_1,\sin\alpha_1),
\end{array}\right.
\eea
that allow us to write
\bea
{\bf z} \,= \,\tilde z^k\ {\bf l}_{(k)}\,,
\eea
so that 
\bea
\frac{\partial z^h}{\partial \tilde z^k} = l^h_{(k)},
\label{proj}
\eea
is the projection operator to the brane $\Sigma_{k}$. 
The two branes $\Sigma_1$ and $\Sigma_2$ are fixed points of $Z_2$ symmetries
acting on the six dimensional space. We wish to investigate the conditions that 
their tensions must satisfy in order to solve the Einstein equations. 
We will use the junction conditions obtained from the Israel formalism.  
We focus on the brane $\Sigma_1$ and find the extrinsic curvature at its position.  
Recall that the second brane $\Sigma_2$ is also a fixed point of $Z_2$ symmetry. 
To implement this information, we require the metric to be symmetric
under $\tilde{z}^2 \to -\tilde{z}^2$.
The transverse space line element in the new coordinate frame changes from 
\begin{equation}
\delta_{ij} dz^i dz^j = d{\bf z}\cdot d{\bf z},
\end{equation}
 to
\bea
{\bf l}_{(1)} \cdot {\bf l}_{(1)}\ d \tilde z^1 d \tilde z^1
+2\, {\bf l}_{(1)} \cdot {\bf l}_{(2)}\ sgn(\tilde z^2)\ d \tilde z^1 d\tilde z^2
+{\bf l}_{(2)} \cdot {\bf l}_{(2)}\ d\tilde z^2 d\tilde z^2
= \tilde\gamma_{mn}\ d\tilde z^m d\tilde z^n,
\eea
with
\bea\label{mm}
\tilde\gamma_{mn}  
&=&\frac{1}{\sin^2 \alpha }
\left(
\begin{array}{cc}
1 & \cos \alpha \,
sgn({\tilde{z}}^2)\\ \\
\cos \alpha \,
sgn({\tilde{z}}^2) & 1
\end{array}
\right)~.
\eea
The bulk line element reads
\footnote{ 
We emphasize  that we are imposing  a
$Z_2$ symmetry at the point
 $\tilde{z}_2=0$, while
at this level we are not imposing a
similar  reflection
symmetry at $\tilde{z}_1=0$. This last
symmetry will be applied  later at the
end of this section,  when 
using Israel  formalism to find the junction
conditions.
}

\bea
ds^2 &=& A^2(t,\tilde z^1,|\tilde z^2|)
\biggl(\eta_{\mu\nu} dx^\mu dx^\nu 
+\tilde\gamma_{mn}\ d\tilde z^m d\tilde z^n\biggr),\\  
A(t,\tilde z^1,|\tilde z^2|) &=& \frac{1}{Ht+{\cal C}_1 \tilde z^1
+{\cal C}_2 |\tilde z^2|},
\\ {\cal C}_1 &=&
\frac{
k_1\cos\alpha_2-k_2\sin\alpha_2
}{\sin \alpha}
~,\quad
 {\cal C}_2 =
\frac{
k_1\cos\alpha_1+k_2\sin\alpha_1
}{\sin \alpha }
~.
\eea


The inverse metric is given by
\bea
g^{MN} &=& A^{-2}\left(
\begin{array}{c|c}
\eta^{\mu\nu} & 0\\ \hline
 0 & \tilde\gamma^{mn}
\end{array}
\right),
\eea
where
\bea
\tilde \gamma^{mn} = 
\frac{\sin^2\alpha}{1-\cos^2\alpha\ sgn^2({\tilde{z}}^2)}
\left(
\begin{array}{cc}
 1& - \cos \alpha\,
 { sgn({\tilde{z}}^2)}\\ \\
 - \cos \alpha sgn({\tilde{z}}^2) \,
 & 1
\end{array}
\right)~.
\eea

In the new coordinate frame the normal vector to the brane $\Sigma_1$  is 
\bea
\tilde n^{(1)}_k = \frac{\partial z^h}{\partial \tilde z^k} n^{(1)}_h = {\bf l}_{(k)}\cdot
{\bf n}^{(1)} = \delta_k^1 =(1,0)\,.
\eea
The normal vector that has a unit-length with respect to the bulk metric is 
\bea
\tilde n^{(1)}_M = A\, {\cal K}\, (0_\mu,1,0)~,
\eea
with 
\bea
{\cal K} = \frac{\sqrt{1-\cos^2\alpha\, sgn^2(\tilde z^2)}}{\sin\alpha}~.
\eea
The projection operator (\ref{proj}) allows us to define the induced
metric on the brane $\Sigma_1$, whose $(\tilde z_2,\tilde z_2)$
component reads~\footnote{The induced  
metric is invariant under bulk diffeomorphisms. Using the bulk frame $(\tilde z_1, 
\tilde z_2)$ one simply gets $g^{(5)}_{\tilde{z}_2 \tilde{z}_2 }
=\delta_{\tilde{z}_2}^m\delta_{\tilde{z}_2}^n A^2 \gamma_{mn} = A^2
\gamma_{\tilde{z}_2\tilde{z}_2}$ that obviously agrees with the result 
obtained in the original $(z_1, z_2)$ frame.} 
\bea
g^{(5)}_{\tilde{z}_2 \tilde{z}_2 }\,=\,l_{(\tilde{z}_2)}^k
l_{(\tilde{z}_2)}^h\, A^2
\delta_{k h} = A^2 ({\bf l}_{(2)})^2 = \frac{A^2}{\sin^2 \alpha}.
\eea
Hence the line element on the brane $\Sigma_1$ is
\bea
\label{metric-1}
ds^{2}_{\Sigma_1} &=& A^2(t,0,|\tilde{z}^2|)\lx(\eta_{\mu\nu}
 dx^\mu dx^\nu 
+
\frac{d\tilde z_2 d\tilde z_2 }{\sin^2 \alpha}\rx).
\eea
Each induced brane metric corresponds to a maximally symmetric space with a 
constant Hubble parameter. For the brane $\Sigma_1$, the Hubble parameter is 
given by
 \bea
 H_1^2 &=&H^2 - 
\sin^{2}{\alpha} \;
 {\cal C}_2^2\,,
 \nonumber
 \label{indmetrs}
 \eea
from which
\be
 \sin{\alpha} \; 
 {\cal C}_2\,=\, -\epsilon_1\,\sqrt{H^2 - H_1^2 },
\ee
where $\epsilon_1$ is equal to plus or minus one and distinguishes
two different branches of solutions. An analogous relation connects 
$H_2$ to ${\cal C}_1$. It is now straightforward to compute the 
components of the extrinsic curvature. Then, we can apply the formalism
of the Israel junction conditions to extract information about the energy 
momentum tensor on the brane. Particularly interesting is
its $(\tilde{z}_2,\tilde{z}_2)$ component that reads
\bea
K_{\tilde{z}_2 \tilde{z}_2}\,=\, \tilde\nabla_{\tilde z_2} \tilde n_{\tilde z_2}
\,=\,-g_{\tilde{z}_2 \tilde{z}_2}
\biggl({\cal C}_1 -\cos \alpha \; {\cal C}_2\biggr)
-A\, \frac{\cotan \alpha}{\sqrt{1-
\cos^2\alpha\, sgn^2(\tilde z^2)}}\, \frac{d}{d \tilde z^2}\, sgn(\tilde z^2)~.
\eea
Defining $w_2 \equiv \tilde z_2/\sin(\alpha_1+\alpha_2)$ in order to remove
the $\sin(\alpha_1+\alpha_2)$ factor from the metric (\ref{metric-1}), one obtains
\bea
\label{K-w2-w2}
K_{w_2}{}^{w_2} =-\biggl({\cal C}_1 -\cos \alpha \; {\cal C}_2\biggr)
-A\, \frac{\cos \alpha}{\sqrt{1-
\cos^2\alpha\, sgn^2(w^2)}}\, \frac{d}{d w^2}\, sgn(w^2)~,
\eea
where we take into account the Jacobean in the delta function $\delta(\tilde
z_2) = \delta(w_2)/\sin \alpha$.

It is important to realise that {\it only} this component of the extrinsic 
curvature contains a singular term $\delta(w_2)$, 
so that $K_a{}^b -\delta_a{}^b K$ has 
singular contribution only along the four dimensional coordinates that 
characterise the intersection. It must be compensated by the energy momentum 
tensor of matter localised at the intersection itself. The maximally symmetric
metric at the intersection reads
\be
ds^{2\hskip0.1cm int}_4 \,=\, A^2(t,0,0)\,\eta_{\mu\nu} dx^\mu dx^\nu, 
\ee 
with Hubble parameter $H$.

The Israel junction conditions relate the components of the extrinsic 
curvature to the energy momentum tensor on the branes. 
A straightforward calculation using the relation between the ${\cal C}_i$
functions and the induced Hubble parameters leads to the following 
conditions


\bea
\Lambda_1 &=& 6M_{5,1}^3 H_1^2
- \frac{8 M_6^4}{\sin \alpha}
\biggl(\epsilon_2 \,\sqrt{H^2-H_2^2}
 -\epsilon_1\,\cos \alpha \sqrt{H^2-H_1^2} \biggr),
\label{condit1}
\\
\Lambda_2 &=& 6M_{5,2}^3 H_2^2
- \frac{8 M_6^4}{\sin\alpha}
\biggl(\epsilon_1 \,\sqrt{H^2-H_1^2}
 -\epsilon_2 \,\cos \alpha \sqrt{H^2-H_2^2} \biggr),
\label{condit2}
\eea
for the Friedmann equations on codimension one branes. In the
intersection we need to carefully determine the contribution from the
singular part of the 6d extrinsic curvature namely
\bea
\left. 2M_6^2\Big(K_\mu{}^\nu-\delta_\mu^\nu K\Big)
\right|_{sing}=-\left. 2M_6^2K_{w^2}{}^{w^2}
\right|_{sing}\delta_\mu^\nu=\frac{\delta\lambda}{2A} \frac{d}{d w^2}\, sgn(w^2)\,\delta_\mu^\nu
\eea
so that making use of expression (\ref{K-w2-w2}) and integrating both
sieds on an infinitesimal interval on gets
 \bea
\delta \lambda = 4M_6^4 \left(\frac\pi 2-\alpha\right)
\eea
 and finally
\bea 
\lambda &=& 3 M_4^2 H^2 - 6 
\biggl(\epsilon_1 \,M_{5,2}^3 \sqrt{H^2-H_1^2}+
\epsilon_2 \,M_{5,1}^3 \sqrt{H^2-H_2^2} \biggr)
+ 4 M_6^4\,\left(\frac\pi 2 -\alpha\right)~.\label{condit3}
\eea
It is important to recognise that the induced gravity terms 
on the codimension one branes provide contributions that are similar to the 
ones that appear in the DGP brane-worlds in five dimensions. 
At the intersection, we find that localised gravity terms on the codimension
one branes induce contributions proportional to $M_{5,i}$ which are again similar
to the ones in the DGP models in five dimensions. Note that by allowing 
the different 5d Newton constant on different codimension one branes, 
the parts of the Friedmann equation proportional to $M_{5,i}$ reproduce  
the asymmetric 5d model with induced gravity considered in 
Ref.~{\cite{asymmetric}}. In addition, we find a term inherited from the six 
dimensional bulk that vanishes in the limit in which the branes form
a right angle \cite{origami}. 

The angle between the branes can be determined by (\ref{cobulk}). 
In terms of the $H$'s and the $\alpha$ it becomes
\be
\Lambda_B = 
-\frac{10}{\sin^2  \alpha}
\Biggl[ 
H^2\left(1+ \cos^2 \alpha \right)-
H_1^2  -H_2^2 
-2 \,\epsilon_1 \epsilon_2\,\cos \alpha
 \,\sqrt{H^2-H_1^2} \sqrt{H^2-H_2^2} \Biggr], 
\label{flatBgen}
\ee
that can be easily solved in terms of the angles:
\be
 \cos \alpha
 \,=\,\frac{\epsilon_1 \epsilon_2\,\sqrt{\left(H^2
 -H_1^2 \right)\left( 
 H^2
 -H_2^2 \right)}-\sqrt{\left(H_1^2 -\frac{\Lambda_B}{10}\right)\left( 
 H_2^2 -\frac{\Lambda_B}{10}\right)
 }}{H^2-\frac{\Lambda_B}{10}}.
\ee

\section{Self-accelerating configurations}
Potentially interesting configurations in our system are 
self-accelerating solutions. It is well known that induced gravity terms 
may provide accelerating cosmological solutions in which acceleration is induced 
by gravity itself and is independent of the presence of matter or cosmological 
constant in the system. This very important observation
\cite{deffayetcosmology,deffayetdvali} has received much attention, since it 
can provide a  model for dark energy
without cosmological constant.
  However, it has also been realized
that this scenario has a serious drawback. The most interesting branch of solutions
in the standard DGP brane-world (the one that contains the self-accelerating 
configuration) is plagued by ghosts~\cite{Luty:2003vm}. It would be interesting to 
study how generic this conclusion is in higher codimensional
models. In order to investigate this issue,  
we start from seeking new self-accelerating solutions in higher
codimensions. 

\subsection{New self-accelerating solutions in 6d spacetime} 
We consider now self-accelerating solutions for our
system. Let us then choose $\Lambda_1=\Lambda_2=\lambda=0$, while the
induced gravity terms (parametrised by $M_4$ and $M_{5,i}$)
do not vanish. For simplicity we assume $M_{5,1}=M_{5,2}$. 
It is then possible to solve our system of equations 
(\ref{condit1})-(\ref{flatBgen}). Taking the difference between 
(\ref{condit1}) and (\ref{condit2}) one obtains
\be
\left(\epsilon_1 \sqrt{H^2-H_1^2}
- \epsilon_2 \sqrt{H^2-H_2^2}\right) 
\left[ 
 \frac{1+\cos{\alpha}}{\sin{\alpha}}
-
\frac{6 M_5^3}{8 M_6^4}\left(\epsilon_1 \sqrt{H^2-H_1^2}
+ \epsilon_2 \sqrt{H^2-H_2^2} \right)
\right]=0,\label{diffC}
\ee
so that we have two possible types of solutions that we now describe.

(i) {\it Symmetric solution}: $H_1=H_2$ and $
\epsilon_1=\epsilon_2$. 
The flat-bulk condition~(\ref{flatBgen}) yields
\be
H_1^2\,=\,\frac{H^2}{2}\,\left( 
1 -\cos{\alpha}
\right).
\label{C}
\ee
From the codimension one equation (\ref{condit1}) one thus gets either the trivial
solution $H=0$, or 
\be\label{forposH}
H= 
\frac{8\epsilon_1}{3 \sin{\alpha}} 
\frac{M_6^4}{M_5^3}
\sqrt{\frac{1+\cos{\alpha}}{2}},
\ee
which, along with (\ref{C}), can be plugged into the intersection
equation~(\ref{condit3}) to give
\be
\frac{1+\cos{\alpha}}{\sin{\alpha}}-\frac14 \left(\frac\pi 2-\alpha\right)
=\,\frac{2\,M_6^4\, M_4^2\,
\left(1+\cos{\alpha}\right)}{3 \, M_5^6\,\sin^2{\alpha}}, 
\ee 
that sets the value for the angle between the codimension-one branes. 
Notice that since the physical range for the angle is $0<\alpha<\pi$, then 
in order to have a positive $H$, eq. (\ref{forposH}) requires a positive $\epsilon_1$.  
 \\[3mm]
(ii) {\it Asymmetric solution}: $H_1\neq H_2$. Defining
\be
{\cal D} = \frac{6M_5^3\sin\a}{8 M_6^4 }\left(
\epsilon_1 \sqrt{H^2-H_1^2}+\epsilon_2 \sqrt{H^2-H_2^2}\right)\,,
\ee 
we get 
\be
{\cal D}\, =1+\cos\a, 
\ee
from (\ref{diffC}). Then the sum of (\ref{condit1})
and (\ref{condit2}), and (\ref{condit3}) give 
\bea
\sqrt{H_1^2+H_2^2} &=& \frac{4M_6^4}{3M_5^3},\\
H^2 &=& \frac{4 M_6^4}{3
  M_4^2}\Biggl[2\frac{1+\cos\a}{\sin\a}-\left(\frac\pi 2 -\a\right)\Biggr]~,
\eea
respectively. Finally from the condition (\ref{flatBgen}) one gets a
relation that sets the value of the angle $\alpha$
and from the positivity condition on $\alpha$
we learn that in this case at least one of the $\epsilon$'s must
be positive as well.


\subsection{Scales of gravity}
A simple manipulation of the previous formulae shows that, in both cases, 
the resulting Hubble parameter is of the order
\begin{equation}
H \simeq \frac{M_{5}^3}{M_{4}^2},
\end{equation}
that is, the same result as in the five dimensional DGP models. 
In order to explain the late time acceleration $H\sim 10^{-42}$ GeV, 
we should require $M_5 =10^{-2}$ GeV with $M_4 =10^{18}$ GeV. 
For $\alpha \sim \pi/2$, the six dimensional Newton constant is roughly given by 
\begin{equation}
M_6^4 \simeq \frac{M_5^6}{M_4^2} \simeq M_4^2 H^2.
\end{equation}
Thus $M_6$ must be $M_6 =10^{-3}$ eV to explain the late time acceleration. 
With these parameters we find that the cross-over scales are given by
\begin{equation}
r_{c,5} \sim r_{c,6} \sim H^{-1}, \quad r_{c,5} = \frac{M_4^2}{M_5^3}, 
\quad r_{c,6} =\frac{M_4}{M_6^2}.
\end{equation}
We should note that these scales are exactly the see saw scale of 
\cite{see-saw}. For large distance, $r> r_{c,5}, r_{c,6}$, 
gravity is five-dimensional or six-dimensional. The laws of 
four-dimensional gravity are valid all the way down to the 
distances of the order $M_{6}^{-1}$. Below this length scale, 
the effective theory of gravity breaks down. As a result, there 
is a lower bound on the scale $M_6$ which comes from 
accelerators, astroparticle and cosmological data, that is 
$M_6 > 10^{-3}$ eV \cite{scales}. This is exactly the scale we need to explain
the present day acceleration of the Universe. Note that for 
a small angle $\alpha \ll 1$, $M_6$ is given by 
$M_6^4 =\alpha M_5^6/M_4^2$. Hence the constraint becomes 
difficult to satisfy. 

However, one realizes that, in order to have $H$ positive, we must choose at 
least one of the $\epsilon_i$ with the positive sign. Then, we are in the same 
class of self-accelerating solutions of \cite{deffayetcosmology}, that are 
notoriously plagued by ghosts. It is then likely that our codimension one branes 
contain ghost excitations. There is a suggestion that if there is no matter on 
the self-accelerating universe, it is possible to quantise the theory without 
having the ghost instability. In our model, we do not need matter on codimension 
one brane. It would be interesting to study at what level the ghost couples 
with matter on the brane at the intersection.

\subsection{More self-accelerating solutions in higher
codimensions}\label{exthc}

In this section we show that the methods previously described
can be easily generalised to a system of $N$ codimension
one branes that intersect in a $(4+N)$-dimensional space-time,
along the lines of \cite{ignaciomulti}.
Their common intersection - a four dimensional, codimension-$N$ 
brane - should correspond to the space-time we observe. We 
consider an Einstein-Hilbert action in the bulk, 
with the addition of a bulk Lagrangian $L$, and we allow for induced 
gravity terms on all the branes, and at the intersections.

The action for the system is then the following
\be
\label{modelg}
\begin{split}
&S=\int_{\rm bulk} d^{4+N} x \sqrt{-g}\ \biggl(\frac{M_{4+N}^{2+N}}2 R+L\biggr)
+\sum_{i=1}^N \int_{\Sigma_i} d^{3+N} x \sqrt{-g_{(i)}}\ \biggl(\frac{M_{3+N}^{1+N}}2
R_{(i)}+L_{(i)}\biggr)\\    
&+\sum_{i\neq j} \int_{\Sigma_i\cap\Sigma_j} d^{2+N} x
\sqrt{-g_{(ij)}}\ \biggl(\frac{M_{2+N}^{N}}2 
R_{(ij)}+L_{(ij)}\biggr)+\cdots
+\int_{\cap} d^{4} x \sqrt{-g_\cap}\ \biggl(\frac{M_{4}^{2}}2 R_\cap+L_\cap\biggr),
\end{split}
\ee
where $\cap\equiv\bigcap_i \Sigma_i$ denotes the three-brane
at the intersection  between 
 all codimension-one branes.
Here we concentrate on the case where the Lagrangians are simply
tensions and the codimension-one branes intersect to form right
angles.  We also describe an empty ``self-accelerated'' scenario where
all the branes undergo a de Sitter phase with no cosmological constant nor
tensions in the system. 
%
%
The bulk geometry is again a maximally-symmetric one $ds^2 =
A^2(t,z_1,\dots,z_N)\ \eta_{MN} dx^M dx^N$ with  
$ A(t,z_1,\dots,z_N) = 1/Ht+\sum_{i=1}^Nk_i |z_i|$ 
and each  static codimension-one brane $\Sigma_i$  is defined by the subspace $\{z_i=0\}$.  
The bulk Einstein equation gives the relation 
\begin{equation}
H^2-\sum_{i=1}^N
k_i^2 = \frac{2}{(2+N)(3+N)}\Lambda,
\end{equation}
whereas the equation of motion for the codimension-one branes
$\Sigma_i$ can be computed by means of the Israel junction condition.
Assuming $Z_2$ symmetry across the branes
\be
\left. {}^{(i)}K^m{}_n =-\frac1{2 M_{4+N}^{2+N}} \biggl(
     {}^{(i)}T^m{}_n-\frac1{2+N}\delta^m_n\ 
{}^{(i)}T\biggr)\right |_{\rm smooth},
\label{junctionN}
\ee
where ${}^{(i)}T^m{}_n$ includes a regular ``matter'' part
$-\Lambda_{(i)}\delta^m_n$ and a part from the smooth part of the
induced EH term, namely  
$-M_{3+N}^{1+N}\ {}^{(i)}G^m{}_n =M_{3+N}^{1+N}\frac{(1+N)(2+N)}{2}\left(H^2-\sum_{i'}
k_{i'}^2\right)\delta^m_n$. Hence
\be
\begin{split}
&\left. {}^{(i)}T^m{}_n\right|_{\rm smooth} =
\Biggl(-\Lambda_{(i)}
 +M_{3+N}^{1+N}\frac{(1+N)(2+N)}{2}\left(H^2-\sum_{i'} 
k_{i'}^2\right)\Biggr)\delta^m_n. 
\end{split}
\ee
Taking the normal vector to the brane to be pointing into the bulk $n_M^+
= A \delta_M^{z_i}$ we get
$
{}^{(i)}K^m{}_n =g^{mr} \nabla_r n_n=\left. \delta^m_n A^{-2}\partial_i
A\right|_{\Sigma_i} = -\delta^m_n k_i.
$
Then Eq.~(\ref{junctionN}) gives
\be
\begin{split}
&\Lambda_i =  M_{3+N}^{1+N}\frac{(1+N)(2+N)}{2}\left(H^2-\sum_{\ell \neq i}  
k_{\ell}^2\right)  +  2(2+N) M_{4+N}^{2+N}k_i ~,
\label{EoM-i}
\end{split}
\ee
which in the special case $N=2$ simply reduces to~(\ref{condit1}) for
a right angle between the branes. The process can be iterated. In fact
each brane $\Sigma_i$ is intersected by 
other $N-1$ branes. From the point of view of an observer sitting on $\Sigma_i$
each of these intersections separate the brane itself in two half parts whose
boundaries are given by that intersection. Consider for definiteness the
intersection $\Sigma_i\cap\Sigma_j$ we will then have a  contribution to the
singular part of the Einstein tensors associated to $R_{(i)}$ 
localised at $z_j =0$ and a contribution to the
singular part of the Einstein tensors associated to $R_{(j)}$ 
localised at $z_i =0$. Such contributions can be evaluated by computing the
extrinsic curvatures associated to the intersection $\Sigma_i\cap\Sigma_j$ as
measured from $\Sigma_i$ and $\Sigma_j$ point of view. Namely,
\be
{}^{(ij)} K^m{}_n +{}^{(ji)} K^m{}_n  = -\delta^m_n(k_i+k_j),
\ee      
and (\ref{EoM-i}) thus generalises as
\be
\begin{split}
&\Lambda_{(ij)} =M_{2+N}^{N}\frac{N(1+N)}{2}\left(H^2-\sum_{l\neq i,j}  
k_{l}^2\right)  +  2(1+N) M_{3+N}^{1+N}(k_i+k_j),
\end{split}
\ee
and so on. In particular for the 3-brane intersection of all the
codimension-one branes we have $N$ contributions from the induced EH terms
localised on the $N-1$-codimension $N$ intersections. Hence 
\be
\Lambda_{\cap} =3 M_{4}^{2} H^2+ 6 M_{5}^{3}\sum_{i=1}^N k_i~.
\label{EoM-cap}
\ee
It is thus clear from the previous setup that the presence of induced EH
terms on the branes allows for localised matter on the branes themselves, at
least in the form of tension. 
%

We then concentrate on a special case of the previous results that gives rise to a de
Sitter solution in absence of any form of matter in the system, namely a
``self-accelerated'' solution. We set $k_i =-H_i$: the bulk equation of motion simply 
 requires $H^2=\sum_{i} H_i^2$, whereas from the junctions conditions (\ref{EoM-i}) one obtains  
\be
\begin{split}
M_{3+N}^{1+N}\frac{(1+N)}{4}\left(H^2-\sum_{\ell \neq i}  
H_{\ell}^2\right) = M_{4+N}^{2+N}H_i ~,
\label{EoM-self-i}
\end{split}
\ee
that, using the bulk equation of motion, reduces to 
\be
H_i = \frac{M_{3+N}^{1+N}}{M_{4+N}^{2+N}}\ \frac{1+N}4\ H_i^2
\ee
which fixes 
\be
K\equiv H_i =\frac4{1+N}\ \frac{M_{4+N}^{2+N}}{M_{3+N}^{1+N}} ,\quad \forall i~.
\label{K}
\ee
The process can again be iterated all the way down to
codimension-$N$. At the last level, we have $N$
contributions to the extrinsic curvature on the 3-brane,  
from $N$ possible intersections of $N-1$ codimension-one branes. Using (\ref{EoM-cap},\ref{K})
and the bulk equation of motion one obtains the 4d Hubble scale and a
relation between all the masses in the system 
\bea
H \!\!\!\!&=&\!\!\!\! \sqrt{N} K = \sqrt{N}~2\frac{M_5^3}{M_4^2},    
\label{4d:dS}\\
2~\frac{M_5^3}{M_4^2}\!\!\!\!&=&\!\!\!
\frac{4}3\ \frac{M_6^4}{M_5^3} 
=\frac{M_7^{5}}{M_6^4}=\cdots= \frac{4}{1+N}\
\frac{M_{4+N}^{2+N}}{M_{3+N}^{1+N}}~.
\label{scales}  
\eea
Let us also stress that all the branes (codimension-one and intersections)
undergo a de Sitter phase. As to the codimension-one branes $\Sigma_i$ the rate is 
simply $H_i = K$, and similarly for $\Sigma_i\cap\Sigma_j$ one has $H_{ij}
=\sqrt{H_i^2+H_j^2} =\sqrt2 K$: for a generic codimension-$n$ intersection, one
obtains $H_{(n)} = \sqrt{n} K$. 

Note that if we require the de Sitter rate $H$ to be equal to today's
Hubble rate $H_0 \sim 10^{-42}~GeV$, from~(\ref{4d:dS},\ref{scales}),
and using $M_4 \sim 10^{18}~GeV$,  we get 
the following values for the mass scales involved in the system
\bea
M_5 \sim 10^{-2} ~GeV, \quad M_6\sim M_5 \left(\frac{M_5}{M_4}\right)^{1/2}\sim 10^{-12}~GeV,
\quad M_7\sim M_6 \left(\frac{M_6}{M_5}\right)^{3/5} \sim 10^{-18} ~GeV \quad \cdots,
\eea  
Hence the higher is the codimension, the lower the fundamental scale
must be. This requirement contradicts to the constraints on the 
fundamental scale $M_{4 +N} < 10^{-3}$ eV and it is impossible to realize 
the self-accelerated scenarios in higher codimensional 
spacetime with $N >2$.

\section{Selftuning configurations}\label{selfsection}

The system of equations  (\ref{condit1})-(\ref{flatBgen})
relates the bulk cosmological constant and brane tensions with the
brane Hubble parameters and the angle between the branes.
Since we have induced gravity terms on the branes, it is possible that 
they are enough to ensure that four dimensional gravity is obtained in the 
relevant regimes at the intersection, with no need to compactify the 
extra-dimensions. Then, the angle between  the branes is a free parameter; 
it is natural to ask whether the {\it selftuning} solutions exist where, by 
changing the tension at the intersection $\lambda$,
only the angle $\alpha$ between the branes changes and 
all the other quantities are fixed.

\smallskip


The following is an example of such solutions
 \bea
 \Lambda_1 &=& 6 M_{5,1}^3 H_1^2, \\
 \Lambda_2 &=& 6 M_{5,2}^3 H_2^2, \\
 H^2 &=& H_1^2  \,=\,  H_2^2 \,  = \, \frac{\Lambda_{B}}{10}, \\
\frac\pi 2 - \alpha  &=& \frac{\lambda-3
   M_4^2 H^2 }{4 M_6^4}.
 \eea
The Hubble parameter at the intersection is {\it independent} 
of the tension $\lambda$. The tension only controls the angle between the 
branes. Then, if we localise standard model physics at the intersection, any contribution
to the four dimensional cosmological constant, that is, the brane
tension $\lambda$, does not curve the four dimensional spacetime. 
Instead, it only changes a parameter in the higher dimensional 
spacetime that is the angle between the branes, realising
the desired
 {\it selftuning mechanism}.

\smallskip

It is important to point out the similarity and difference 
between this idea and the selftuning scenarios in six dimensions using 
braneworlds on conical singularities \cite{carroll,burgess1} where two extra 
dimensions are compactified by a magnetic monopole. In these cases, the selftuning 
mechanism suffers from a quantisation condition for the magnetic monopole responsible 
for compactifying the extra space and we need a fine-tuning between the 
cosmological constant and other parameters of the models. Our case is different 
because in principle we do {\it not need} to compactify the extra dimensions.   
In our model, the induced gravity is supposed to recover 4D behaviours 
of gravity on scales smaller than the cross-over scales $r_{c,5}$ or $r_{c,6}$.
Similarly to what happens in~\cite{carroll,burgess1} and in their infinite-volume
counterparts~\cite{dvaligabadadze,puretension2} 
(which are more alike to the present configuration), also in this case
the vacuum energy density has a finite critical value 
that corresponds to the maximum of the deficit angle (minimum angle
between the branes): in the limit $\lambda \gg 3M_4^2 H^2$ and using
$\alpha_{crit} =0$,  one gets 
\bea
\lambda_{crit} = 2\pi M_6^4~.
\eea
Although there is a critical value for the brane tension, our solution
is topologically different from a  
vortex due to the existence of codimenison one branes. This would have an 
important implication for the existence of the ghost. 
It has been shown that 
there appears a ghost if we consider a 4d vortex with induced gravity 
in a 6d spacetime depending on the regularisations \cite{6d-induced1}.
We should carefully re-examine this issue in our model, as a mechanism
like the one found in~\cite{newdvali} might be at work and help cancelling ghosts.  

\section{Conclusions and open issues}
In this paper, we studied maximally symmetric solutions for 
intersecting brane-worlds with induced gravity terms.
We found  new classes of the self-accelerating solutions that 
realize  de Sitter solution without any cosmological constant 
in the system. In a 6d spacetime, this realizes the see-saw 
modification of the gravity. The 6d Planck scale must be 
$M_6=10^{-3}$eV to explain the late time acceleration $H =H_0
=10^{-42}$ GeV. This scale determines the IR scale where 4d gravity 
is modified $M_6^2/M_4 \sim H_0$. On the other hand, $M_6$ defines 
the UV scale where 4d gravity is modified. It was shown that 
the lower bound for the fundamental Planck scale is roughly 
$10^{-3}$ eV from various constraints which is consistent 
in the 6d case. The small angle between the two branes 
makes it difficult to satisfy this constraint. We have shown 
that the construction of the self-accelerating universe with 
right angles between the branes can be naturally extended to 
$N$-dimensional spacetime. However, it was shown that, in order to 
realize the observed late time acceleration, the 
higher-dimensional Planck scale is increasingly lower  and violates the 
lower bound for the fundamental scale. We should mention that 
the problem of the ghost instability would exist even in 
these new solutions. It would be important to study at what level 
the ghost couples with matter on the brane at the intersection.

It is also possible to find the self-tuning solution where 
the cosmological constant on the intersection only
depends on the angle between the branes.
 Unlike other examples, in principle 
we do not need to compactify the extra dimensions, since
the induced gravity terms may be sufficient to provide four
dimensional gravity in the relevant regimes. Then, the 
angle between the branes is a free parameter that
does not have to satisfy any constraint. 
In the vortex case,
it is shown that a regularisation is needed to remove the 
divergence of 6d propagator and a ghost can appear depending on 
the regularisation. It is vital to re-examine these issues for 
our intersecting branes.    

\smallskip

Finally, let us briefly discuss the problem of obtaining consistent
four dimensional gravity at the intersection in this set-up. 
The first simple step toward the understanding of this issue is 
to generalise the maximally symmetric solutions to 
the case of FRW cosmological configurations on the branes. Then, by studying 
Israel junction conditions, one can derive the effective cosmological equations 
on the codimension one and codimension two brane-worlds. 
By obtaining the Hubble equation on the codimension two brane-world at the intersection, 
it is possible to determine on which scales standard four dimensional
cosmological expansion can be obtained at the intersection. The resulting cosmological  system presents many interesting features. It is possible to show that the preferred 
energy momentum tensor can be localised at the intersection, although in the case in 
which it is different from pure tension, there is normally an exchange of energy
density between the codimension one branes and the intersection. The details of the 
induced cosmology, as well as their consequences, are going to be analysed in a 
companion paper 
\cite{inpreparation}.

\vskip1.4cm

\subsection*{$\hskip6.8cm$Acknowledgements}

We thank Hyun Min Lee and  Nemanja Kaloper for useful discussions.
We also thank Oriol Pujolas for pointing out an error 
in the identification of the 6D singular structure in the 
earlier version of this paper. 
The work of OC is partly supported by a
Marco Polo fellowship of the University of Bologna.
KK is supported by STFC. GT is supported by 
MEC and FEDER under 
grant FPA2006-05485, by  
CAM under grant HEPHACOS P-ESP-00346, and
by the 'UniverseNet' network  (MRTN-CT-2006-035863).
 OC and GT
 are grateful to
the Institute of Cosmology and Gravitation of the University of
Portsmouth for support and warm hospitality while this work was initiated.

\end{document}